# Residential structure survivability to large wildfires in the United States

Mukesh Kumar[1], John T. Abatzoglou[2]*, Crystal A. Kolden[1], Mojtaba Sadegh[3,4]

[1]Fire Resilience Center, University of California, Merced, Merced, CA, USA
[2]School of Engineering, University of California, Merced, Merced, CA, USA
[3]Department of Civil Engineering, Boise State University, Boise, ID, USA
[4]United Nations University Institute for Water, Environment and Health, Richmond Hill, ON, Canada

*Corresponding author. Email: jabatzoglou@ucmerced.edu

## Abstract

Wildfire impacts on US communities have escalated in recent decades, highlighting the need to better understand factors that influence wildfire outcomes. We find that 567,000 homes were exposed to wildfires across the contiguous US during 2001–2020, two-thirds of which occurred and increased five-fold in the Western US. While residential structure survivability—the percent of structures within a wildfire perimeter that survive the fire—remained stable in the Eastern US in the past two decades, it declined by 10% in the West. Survivability was explained by structural age, surrounding fuels, and fire weather. Survivability was 87% for homes built pre-1990 compared to 92% for post-1990 homes in the West. Survivability was lowest in forests compared to grasslands and shrublands. Finally, survivability was markedly lower for fires coincident with extreme fire weather. Our results suggest that modern building codes, fuel management, and proactive planning can strengthen wildfire resilience.

## 1. Main

Wildfire, hereafter referred to as fire, disasters and associated socioeconomic losses have surged across the United States (US)[1,2] and globally during the past two decades[3,4]. Area burned, large-fire frequency, suppression expenditures and—most critically—losses of homes are now at or near modern record highs, particularly in the western US[5–7]. The number of people exposed to large fires in US more than doubled in the past two decades[8], and western structure losses more than tripled during 1999-2020[2]. Recent analyses further reveal a two-fold increase in home exposure within fire perimeter since 1990s in the US[9], and that the fastest-growing fires account for nearly 90% of all structures damaged or destroyed despite representing only 2.7% of events[10]. These studies provide critical baselines for loss, but they do not quantify the survivability rate of structures directly exposed to fire. While most research and policy remains loss-centric, disaster-risk scholars argue for resilience-centric approaches that emphasize what persists and why[11–13] by quantifying and contextualizing loss relative to survival. Here, we introduce home survivability—the proportion of exposed residential structures that are not destroyed within the footprint of a fire—as a resilience-oriented metric that aligns with emerging approaches in fire management and sustainability science[14,15].

Escalating risk is driven by several intertwined forces, including longer, drier fire seasons with more frequent critical fire-weather events[16,17], wildland-urban interface (WUI) expansion into fire-prone landscapes[6,9,14,18], and unsustainable fuel accumulation as a result of decades of fire suppression in historically fire-frequent forests[19,20]. Home loss is governed by building characteristics, defensible space, neighborhood density[21–25], and extreme fire weather such as downslope winds[26]. Regional assessments have shown that homes built with modern codes survive at higher rates than older buildings[27,28] and that losses vary by vegetation type[2,9,29], but there has been limited assessment of survival rates nationally. Existing national studies either pool all structure classes, focus solely on hazard parameters, such as burned area, or examine individual events, limiting their utility for national-scale forward-looking risk reduction[13,25,30,31]. To address these gaps, we analyze two decades (2001–2020) of incident-level fires, and associated housing age, vegetation, and fire-weather data for the US. Our objectives are to: (i) quantify spatial and temporal patterns of home survivability, (ii) determine how structural age, surrounding vegetation, and extreme fire weather influence exposure and survival, and (iii) investigate regional drivers of home survivability to inform adaptive, evidence-based strategies for future fire risk reduction. By shifting the lens from loss to survival and linking multiscale biophysical and socio-structural data, our study provides a much needed national benchmark for how many homes withstand fires—and why—under the interacting pressures of climate change and WUI growth.

## 2. Results

### 2.1. Patterns and trends of residential structure exposure and survivability to fires

Exposure and survivability to large fires are quantified using incident-specific residential structure loss records from the ICS-209-PLUS dataset[32] where such losses occurred within the large fire perimeters from Monitoring Trends in Burn Severity[33] (MTBS). We considered all large fires that had at least one exposed residential home. We find that during 2001–2020, the western US (11 westernmost states in the contiguous US, also referred to as West herein; Fig. 1a) cumulatively accounted for 380,130 home exposures (67% of the national total) to large fires (n=4,225)and the Eastern US large fires (n=7,342) accounted for 186,492 home exposures (33%). Annual home exposures increased fivefold in two decades in the West (p = 0.054). The trend in western exposures aligns with broader evidence of expanding burned area and WUI growth under warming and drying conditions[6,17,34] and rising population exposure to fire[8]. In contrast, home exposure in the East (Fig. 1b) remained comparatively stable during 2001–2020. Average per fire home exposure in the East (25 homes per fire) was also notably lower than in the West (90 homes per fire), reflecting the separation between wildfire footprints and human settlements in the East.

California accounts for approximately 46% of all homes exposed to fire across the US, despite representing only ~6% of fire occurrences and ~10% of homes (Fig. 1c, Fig. S1a, Table S1). Moreover, California has the highest mean per fire exposure (351 homes per fire) during 2001-2020, 7-fold greater than the national average (49 homes per fire). This concentration underscores the intersection of extensive WUI, rapid development over the latter half of the 20th century, extreme fire weather, and the outsized impact of rare, catastrophic fire events[34] in a Mediterranean biome, associated with the highest rate of wildfire disasters globally[3].

State-level survival rates (Fig. 1d) display a distinct regional divide. Most eastern states exhibit survival rates above 98%, except for Tennessee (68%) and Texas (87%). In these states, losses were heavily influenced by singular extreme loss fires including the 2016 Chimney Tops 2 Fire in Gatlinburg, Tennessee, which destroyed 2,013 homes with a 28% survivability rate, and the 2011 Bastrop County Complex Fire in Texas that destroyed 1,669 homes with a 24% survivability rate. Pooled state-level fire survivability rates were lower in the western US compared to the eastern US, with California (85%) and Colorado (90%) having the lowest rates. In the West, rare but severe incidents similarly exerted a disproportionately large impact on the overall statistics; for example, the 2018 Camp Fire in California destroyed 13,983 homes, making it the most destructive fire on record with the lowest (14%) survivability. Trends in survival rates further highlight regional contrasts. In the West, survival rates exhibit high interannual variability (Fig. 1e), with relatively low survivability in major fire years. Nevertheless, the overall survivability declined 10% during 2001–2020 ($p = 0.064$) in the West (Fig. 1e), with a 15% decline in structure survivability in California ($p = 0.046$) (Fig. S1b). In contrast, survival rates in the East (Fig. 1f) remain consistent (98–100%), without any significant trend ($p = 0.555$) during 2001–2020.

Fire size emerges as a key determinant of survivability in the West. Survival rates decline sharply with increasing fire size in the West (Fig. 1g), dropping from 97% in 1,000–10,000 acres fires to 90% for fires burning 10,000–100,000 acres and further down to 80% in fires exceeding 100,000 acres burned. This pattern likely reflects the compounding challenges of suppression, evacuation, and exposure in larger fires. California shows a similar decreasing pattern but with even lower survivability of 73% for fires >100,000 acres (Fig. S1c). In the East (Fig. 1h), survival rates decline from 98% for fires burning 1,000–10,000 acres to 83% for fires burning 10,000–100,000 acres. However, the survival rate was 96% for 100,000 acres in the East. This divergence may be due to the small sample size ($n = 19$) and their occurrence in remote areas far from communities.

Our analysis further reveals that a small number of "black swan events"—i.e., individual, unusually destructive fires—have disproportionately contributed to the number of homes destroyed by or exposed to fire (Table 1). While there have been several such extreme fires in the West, we highlight only the top two most destructive events here (the 2017 Tubbs Fire and the 2018 Camp Fire). In the East, we also highlight two black swan events with the highest destroyed homes—the 2011 Bastrop County Complex Fire in Texas and the 2016 Chimney Tops 2 Fire in Tennessee (Table 1). Excluding black swan events increases the overall survivability from 85% to 91% in California, 68% to 98% in Tennessee, and 87% to 92% in Texas (Fig. 1d). These four events, which collectively account for only ~6% of total exposed homes in the US (33,983 of 566,622), caused 48% of home losses in the past two decades. Their outsized impact underscores the critical importance of hardening structures and managing fuels to prepare for rare but high-risk extreme fire-weather windows.

**Table 1. Black swan fire events.** Top two most destructive fires, defined by total homes destroyed in the East and West, separately, during 2001 to 2020.

| Year | Black swan events | Survival rate | Homes exposed | Homes destroyed | Key drivers |
|------|-------------------|---------------|---------------|-----------------|-------------|

| 2017 | Tubbs Fire/ Central LNU Complex, CA | 59% | 12,898 | 5,303 | Downslope winds, drought-cured fuels |
| 2018 | Camp Fire, CA | 14% | 16,095 | 13,893 | Downslope winds, drought |
| 2011 | Bastrop County Complex Fire, TX | 24% | 2,186 | 1,669 | Drought, strong winds (from Tropical Storm Lee) |
| 2016 | Chimney Tops 2 Fire, TN | 28% | 2,804 | 2,013 | Strong winds, drought |

## 2.2. Modern construction improves fire survivability in the West

Among the homes exposed to fires in the US, the construction year shows associations both with likelihood of exposure to a large fire and probability of survival (Fig. 2a–i). The houses built before 1950 account for nearly a quarter of all homes exposed to fire (Fig. 2a). Western and Southern states contain comparatively fewer structures of this age group, reflecting later settlement and more recent suburban and WUI growth (Table S1). Homes built between 1950 and 1989 during the post-war building boom account for 40-60% of exposed residences nationwide (Fig. 2b). Exposures of more modern homes built after 1990 occurred less uniformly across much of the US, with distinct hotspots in the intermountain West and Southern states (Fig. 2c). These factors partially reflect geographic differences in housing stock age by state (Table S1) and growth patterns of expansion from urban centers into wildlands.

Modern construction improves the survivability of homes to fire; more so in the western US (Fig. 2d–f; g–i). Western states saw higher survival rates in more recently built homes; 87% survivability in pre-1950 and 1950–1989 builds and 92% for 1990 onward builds (Fig 2g-i). Notably in California, homes built in and after 1990 had a higher survival rate of 89% (8,223 destroyed out of 86,223 exposed) than mid-century (1950–1989) homes, which had a survival rate of 83%, and pre-1950 homes, with the lowest survival rate of 81% (Fig. 2d–f). Eastern states had an overall higher survivability rate (96%) regardless of the built year (Fig 2g-i).

## 2.3. Survivability is higher for structures adjacent to non-forest vegetation in the West

Structure survivability varies by surrounding vegetation type (Fig. 3a–e). In the West, survival rates are lower in forests (83%) than shrublands (93%) and grasslands (96%), showing a combination of more extreme fire behavior, longer suppression times, and the challenges of defending homes in complex fuel mosaics (Fig. 3e). In the East, survival rates are comparatively higher: 96% in forests, 98% in shrublands, and 97% on grasslands. The total number of homes exposed to fire, categorized by vegetation class (Fig. 3d), shows notable regional differences. In the West, substantially more structures are exposed in shrublands compared to the East. This pattern aligns with regional differences in where residential development occurs across fuel types as well as in the WUI[35] and with the tendency for large, fast-moving fires to occur in more open fuels[10], where a high concentration of structures increases total exposure and, consequently, the number of homes destroyed.

The spatial distribution of residential structure exposure across US by primary vegetation class highlights distinct patterns at the state level consistent with land cover patterns, with forests being the dominant surrounding vegetation for residential fire exposure across most of the eastern states (Fig. 3a). Shrubland-associated exposure is concentrated in the intermountain West where shrublands dominate (Fig. 3b). Grassland exposure is most pronounced in the Great Plains (Fig. 3c). See Fig. S3 for complementary analysis of land cover types. The analysis of anomaly in fraction of home exposure and fraction of area burned in each vegetation class also shows marked patterns among states (Figs. 3f-h). Notably, California had relatively more structure exposure in shrublands—likely reflecting the significant WUI in southern California that has been subjected to large loss fire events (Fig. 3f). In the Great Plains, forest home exposure percentage surpasses forest burned area percentage (Fig. 3f-h), whereas in other states the fractions of exposed homes and burned area are comparative across the vegetation classes.

## 2.4. Extreme fire weather reduces home survivability

Most residential structure exposures in the West occurred during severe fire weather (Fig. 4a-f), with nearly half (48%) of exposures during extreme fire weather that included winds (defined by 99$^{th}$ percentile Burning Index, BI) values (Fig. 4e). A similar pattern is observed in the East, where 33% of all exposure occurred during the highest BI days (Fig. 4e). Similar trends are observed for extreme fire weather defined by Energy Release Component (ERC) (Figs. 4a-c). Although both indices reflect fire potential, BI incorporates wind, making it more sensitive to rapid, wind-driven fire events that include substantial ember cast, explaining the greater residential exposure under high BI than under ERC alone.

Residential structure survival rates decline with intensifying fire weather (Fig. 4c, f). In the West, survival rates drop to ~80% during the most extreme fire weather windows (Fig. 4c). Similar patterns are seen in the East (Fig. 4c, f). The West also experiences a greater number of large fires with higher spread rates than the East (Fig. 4a, d).

## 3. Discussion

Our study shows that from 2001–2020, the western US accounted for two-thirds of the 566,622 homes exposed to large fires nationally, with California alone accounting for nearly half of total exposures. The residential structure exposure in the West increased by approximately fivefold from 2001 to 2020. The increasing exposure co-occurred with a nearly 10% decline in survival rates in the West, including a 15% decline in California. In contrast, the East maintained relatively stable exposure levels and consistent survival rates. The southeastern states contributed substantially to the total exposure but maintained comparatively higher survival rates, suggesting that exposure alone is insufficient to predict structural damage[36–38]. Relationships between the structure age and fire survivability demonstrate that modern building codes can mitigate fire risk even under challenging weather conditions. However, the persistence of vulnerable older housing stock—particularly areas of pre-1990 housing stock in regions where climate change is increasing the frequency of extreme fire weather events[28,30,39]—creates a vulnerability to future fires.

Residential structure survivability to fire is consistently higher (≥ 96%) in the East, whereas in the West, survival rates are markedly lower, especially in forests (83%). These patterns reflect not only more extreme fire behavior in the West—such as intense rapid fires and long-distance ember transport[9,40]—but also longer suppression times and the challenge of defending homes in complex forest fuel mosaics[2,23,28]. Intense crown fires, long flames, and ember production in western forests expose adjacent homes to risk[41,42]. Despite our finding of the lowest survivability rate in forests, the largest number of fire-destroyed homes occur in shrublands and grasslands due to higher exposures within these landscape types, particularly in Mediterranean California.

Our fire weather analysis shows that extreme conditions serve as critical catalysts for both increased residential exposure and decreased survival rates. Nearly half of all residential exposures in the West occur during extreme fire weather conditions, for which structure survivability was significantly lower. These patterns highlight that the fire impact is markedly skewed toward short windows of extreme fire weather. Extreme fire weather conditions present a compounding effect, enabling a larger number of successful ignitions and greater fire spread potential[43], resulting in nearly twice as many large fires under most extreme BI conditions compared to moderate BI conditions in the West.

Finally, the outsized impact of the most intense fire weather windows underscores the critical importance of mitigation strategies. The increasing frequency and intensity of extreme fire weather under climate change suggests increased fire exposure and impact in the future[44–46]. Adaptive and transformative management approaches are essential for enhancing resilience under the changing conditions[19,47]. While "black swan"-type fire disasters may be unpredictable, their occurrence is virtually certain without substantial shifts in building and land stewardship, given continued development in fire-prone environments under intensifying fire weather conditions[6,46,48]. The increased probability of catastrophic fires in fire-prone socio-ecological systems in the absence of substantial shifts in perspective and policy has important implications for risk management and decision-making. Rather than designing for average conditions with occasional extreme events, effective strategies must treat extreme conditions as the design standard for infrastructure, building codes, suppression efforts, and community preparedness[49]. Resilience to worsening fire risk will depend on combining modern building standards, coordinated fuel management at the landscape scale, and proactive risk planning in areas vulnerable to severe fire weather[50].

## 4. Methods

This study analyzed fire impacts on homes across the US during 2001–2020. Fire incident records were obtained from the ICS-209-PLUS dataset[32], which includes detailed incident-level data on destroyed residential structures, fire duration, acres burned, and related incident attributes. ICS-209-PLUS incidents were spatially joined with MTBS[33] fire perimeters to estimate the number and attributes of structures located within the perimeter of each large fire. Inclusion criteria for fire events were set to MTBS standards: fires ≥500 acres (202 ha) in the Eastern US and ≥1,000 acres (404 ha) in the Western US were considered, focusing on large fires only. Quality control procedures (see section S1) were applied to eliminate records with implausible or inconsistent values.

Home exposure was calculated by determining how many residential structures occurred within each MTBS fire perimeter ("direct exposure"). This was done using the high-resolution census block group data from the American Community Survey (ACS) 5-year housing estimates[51] from 2013 to 2020 (see section S2). For each fire, the MTBS perimeter was overlaid with census block groups, and if the block group were completely inside the fire perimeter, total number of homes present within those block groups were counted as exposed homes. If a block group partially intersected the fire perimeter, an area-weighted approach was used where we multiplied the fraction of the block group's area inside the perimeter by its total housing count, then summed these across all intersected groups as well as homes that were completely inside fire perimeters to calculate the number of homes exposed ("exposed structures"). All spatial datasets were projected to the North America Albers Equal Area (EPSG:5070) for consistent and accurate area calculation. In addition, exposure was calculated independently for each fire event, meaning that if a structure or block group was exposed in multiple years or events, it was counted in each instance. This method ensured that annual exposure totals represent the sum of event-based risks rather than unique structure counts. To calculate survivability, we first obtained the count of destroyed residential structures for each fire using the "STR_DESTROYED_RES" field from ICS-209-PLUS, which only reflects homes reported as destroyed (not damaged or threatened). We then computed survivability as the percentage of exposed residential structures within the fire perimeter that were not destroyed. This approach provides a robust, spatially consistent, and fire-by-fire calculation of both exposure and survivability for residential homes located within the large fire perimeter.

The ACS housing data from 2013 to 2020 were further categorized by construction year into three broad groups: <1950, 1950–1989, and ≥1990, using table "B25034" which offers information on the year houses were built (section S2). This provided proportions of three different age groups of exposed structures at the census block group level for each fire perimeter. Furthermore, the age group proportionality from exposure was then assigned to the destroyed structures in census block for each fire due to limited geolocated data for each destroyed structure. While this approach involved estimation, it leveraged the finest available spatial resolution—the census block group level. This classification enabled an analysis of how the construction era of residential homes with differences in development patterns, building codes, and materials affect fire exposure and survivability.

Vegetation types at exposure sites were assessed at the pixel level, overlaying housing and destruction record with the 2020 LANDFIRE Biophysical Settings (BPS) geospatial data[52]. For each fire, all pixels intersecting with a fire perimeter were classified by the proportion of vegetation type, i.e., forest, shrubland, and grassland. We found that <1% of total exposures occurred in land cover types which were classified as "Other" as seen in Figure S3. Rather than selecting a single dominant class per fire, our pixel-based approach quantified the number of structures exposed within each vegetation class for each fire. We then calculated summaries of total exposure and survivability associated with each of the three vegetation types both at the state and regional scales.

Fire weather conditions were characterized using gridded daily meteorological data from GridMET[53], focusing on two indices from the US National Fire Danger Rating System: ERC and BI – both for a uniform fuel class (dead conifer with heavy fuels). For each large fire, daily ERC and BI values were extracted from the nearest grid cell to the MTBS fire centroid at the day of impact from ICS 209-PLUS records, and percentiles were calculated relative to the local 2001–

2020 climatology. The "worst weather window" was defined as the maximum value of the fire weather index between the discovery date and the date at which the fire reached 75% of the final fire size. To robustly analyze links between fire weather intensity, structure exposure, and survival rates, fires were binned by the percentile of worst-weather conditions (≤50th, 50–70th, 70–80th, 80–90th, 90–95th, 95–99th, and ≥99th). Finally, we computed the survival rates, fire counts, and total exposed structures within each ERC and BI percentile bin.

Substantial year-to-year variability in fire activity and the limited time span of fire records contribute to uncertainty and anomalies reported herein. Such uncertainty is common in studies that depend on spatially and temporally resolved fire datasets, which only reliably cover recent decades[4,8,54]. Nevertheless, our study provides insight into recent spatial and temporal patterns of home exposure and survivability to large fires across the US.

### Acknowledgements

The work was partially supported through grants from the NSF (OAI-2019762) to J.T.A., the Joint Fire Science Program (21-2-01-3) to J.T.A., M.S., and USHUD (H-221738CA) to C.A.K. and M.K.

### Author Contributions

Mukesh Kumar: Conceptualization, Data curation, Formal analysis, Investigation, Methodology, Resources, Validation, Visualization, Writing – original draft, Writing – review & editing

John T. Abatzoglou: Conceptualization, Funding acquisition, Investigation, Methodology, Supervision, Writing – review & editing

Crystal A. Kolden: Conceptualization, Methodology, Supervision, Writing – review & editing

Mojtaba Sadegh: Conceptualization, Methodology, Resources, Writing – review & editing

### Conflicts of Interest

The authors declare no conflicts of interest relevant to this study.

### Data Availability Statement

The datasets used in this study are openly available, as cited in the reference section.

**Figures**

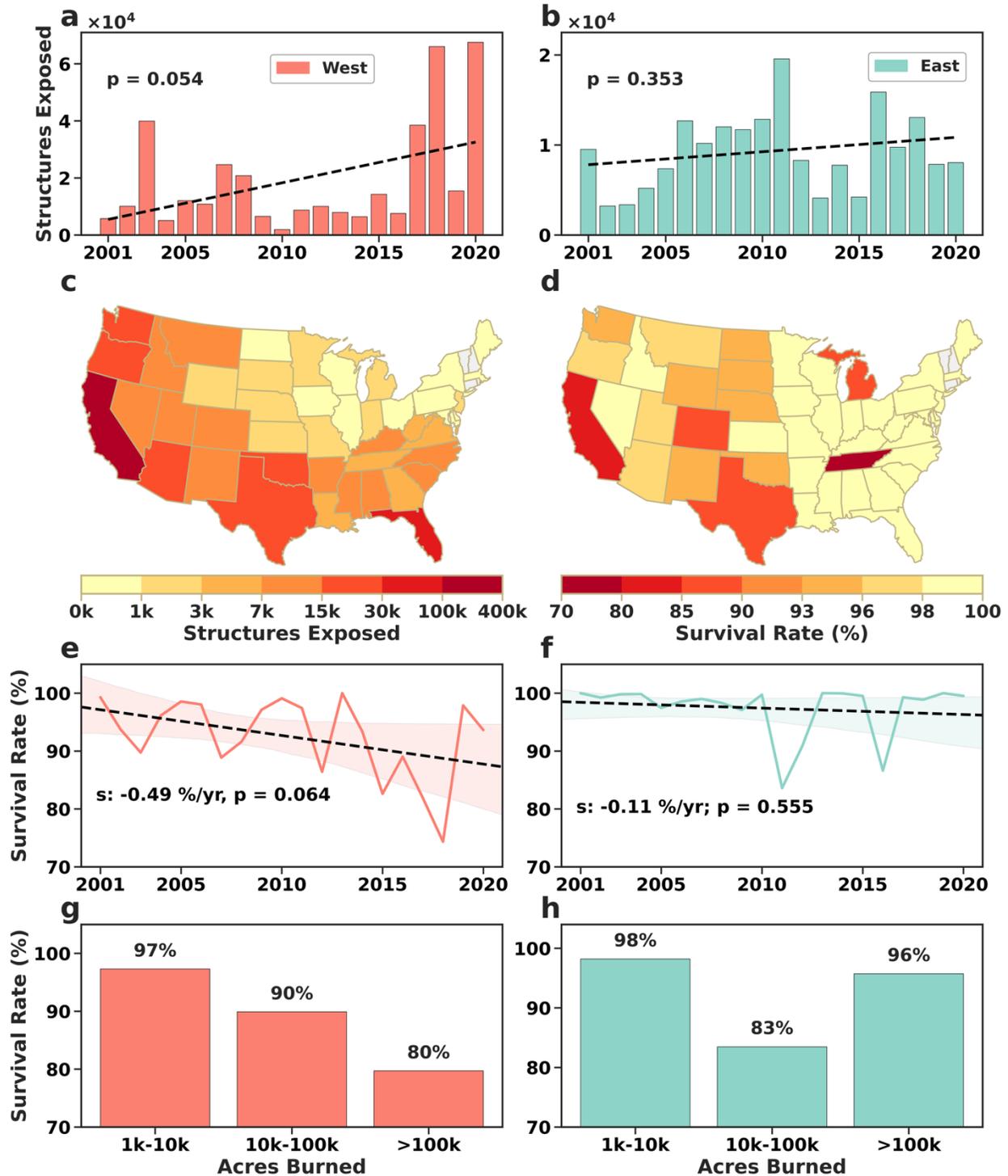

**Figure 1. Exposure and survivability of residential structures to fire in the US.** (a, b) Trends in annual residential structures exposed to fires in the (a) Western and (b) Eastern US. (c) Cumulative exposure of residential structures to fire by state. (d) State-level residential structure survival rates. States shown in gray in (c) and (d) had no large fires during the analysis period. (e, f) Trends in annual residential structure survival rates in the West (e) and East (f). (g, h)

Survival rates by fire size class for the West (g) and East (h). Fires are grouped into three final burned area classes: 1,000–10,000 acres, 10,000–100,000 acres, and greater than 100,000 acres.

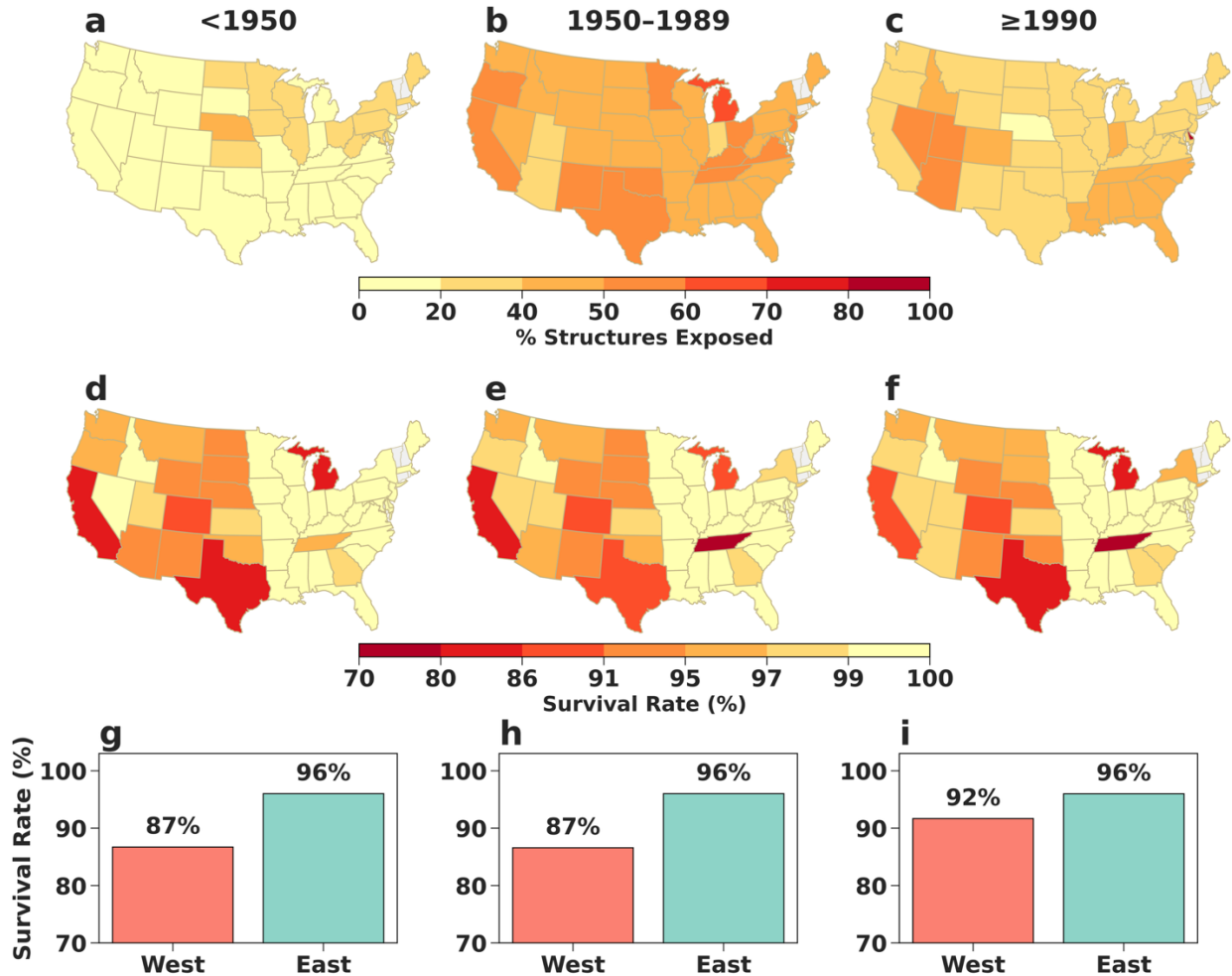

**Figure 2. Relationship between age of residential structures and exposure and survivability to fire.** (a–c) Percentage of structures exposed to large fires in each state built before 1950 (a), between 1950–1989 (b), and in and after 1990 (c) (darker red = higher exposure), calculated as a share of total exposed structures within each state. This figure reflects the age group of exposed structures in each state, not their share of national exposure. (d–f) Corresponding state-level survival rates (percentage of exposed homes not destroyed; lighter yellow = higher survival). States shown in gray in (a-f) had no large fires during the analysis period. (g–i) Survival rates for the West and East by age group.

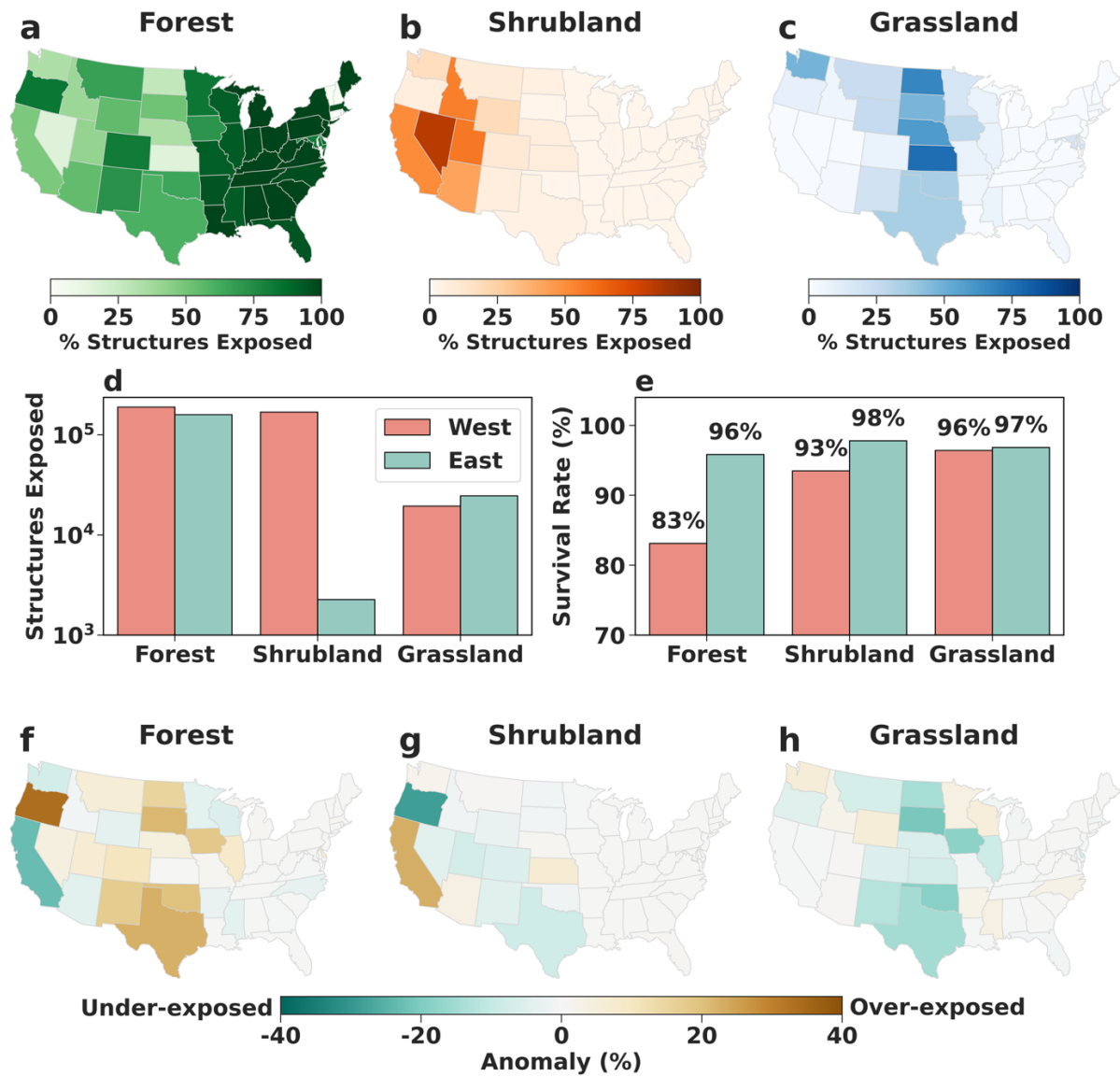

**Figure 3. Vegetation types shape structure survivability to fire.** (a–c) Percentage of residential structures exposed to fire in different types of vegetation, mapped at the state level: forest (a), shrubland (b), and grassland (c). For each state, values represent the proportion of all exposures occurring within each vegetation type. States shown in gray in (a-c) had no large fires during the analysis period. (d) Total number of residential structures exposed to fire by vegetation type and region (West vs. East). (e) Survival rates of exposed residential structures by vegetation type and region. (f–h) Anomaly (%) calculated as the difference between percentage of exposed residential structures and percentage of burned area in each vegetation type: forest (f), shrubland (g), grassland (h). Positive anomalies indicate vegetation types with a higher proportion of homes relative to the burned area.

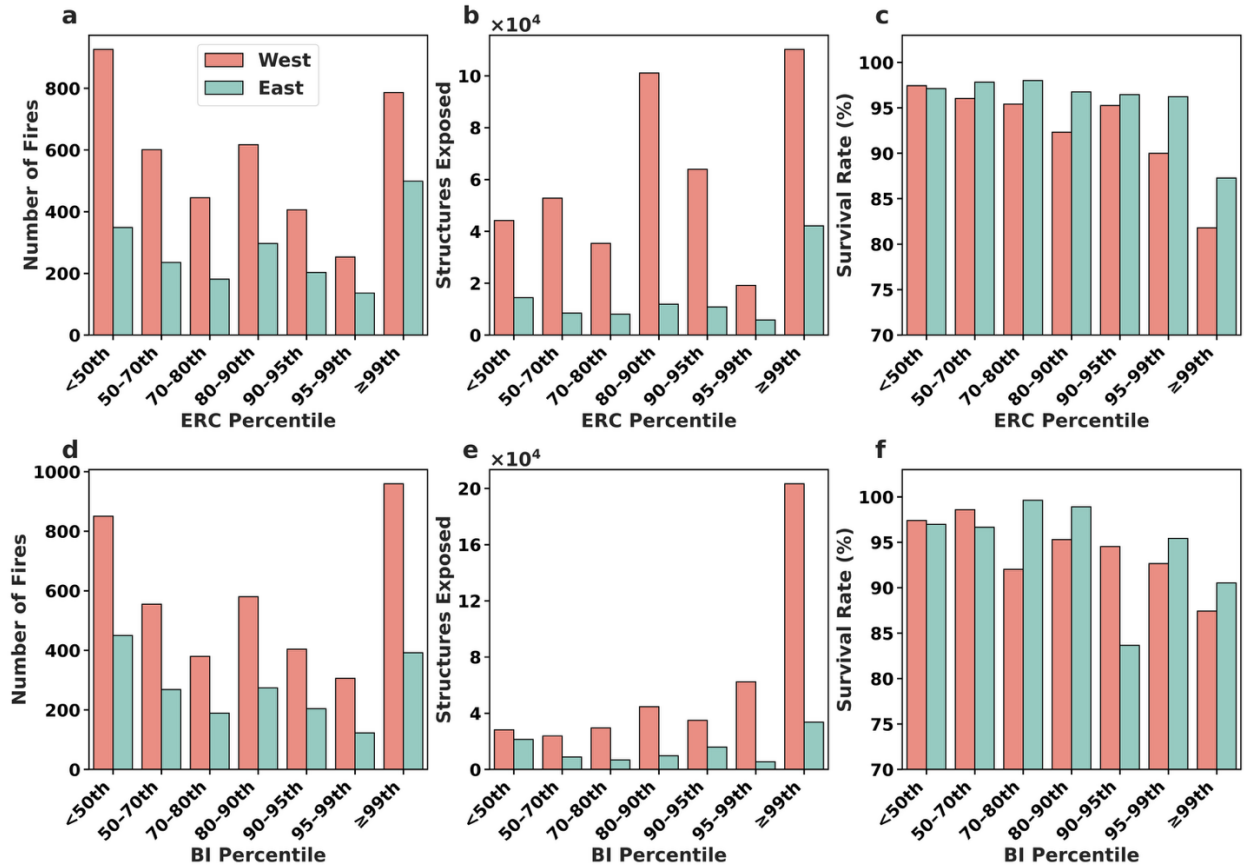

**Figure 4. Extreme fire weather conditions reduce survival rates and increase home exposure.** (a, d) Number of fires within each Energy Release Component (ERC) and Burning Index (BI) percentile bin. (b, e) Total number of residential structures exposed to fire within each ERC and BI percentile bin. (c, f) Survival rates of residential structures exposed to fire, stratified by ERC percentiles and BI percentiles.